\documentstyle[prl,tighten,aps,epsfig,amstex,multicol,palatino]{revtex}
\begin{document}

\newcommand{\ncd}{\newcommand}
\ncd{\ra}{$\rightarrow$}
\ncd{\ds}{\displaystyle}
\ncd{\CNOT}{\mbox{CNOT}}
\ncd{\St}{ $S$ }
\ncd{\Sm}{S}


\title{Quantum computing via measurements only}

\author{Robert Raussendorf and Hans J.\ Briegel}
\address{Theoretische Physik, Ludwig-Maximilians-Universit{\"a}t
M{\"u}nchen, Germany}
\date{\today}
\maketitle

\begin{multicols}{2}
\narrowtext

{\bf 
A quantum computer promises efficient processing of certain computational
tasks that are intractable with classical computer technology 
\cite{bennett-divincenzo-review}. While basic principles of a quantum computer 
have been demonstrated in the laboratory \cite{experiments}, scalability of these 
systems to a large number of qubits \cite{cirac-zoller-2000}, essential for practical 
applications such as the Shor algorithm, represents a formidable challenge. Most of 
the current experiments are designed to implement sequences of highly controlled 
interactions between selected particles (qubits), thereby following models of a 
quantum computer as a (sequential) network of quantum logic gates 
\cite{deutsch89,barenco95}.
Here we propose a different model of a scalable quantum computer. 
In our model, the entire resource for the quantum computation is provided 
initially in form of a specific entangled state (a so-called cluster state) 
of a large number of qubits. Information is then written onto the 
cluster, processed, and read out form the cluster by one-particle 
measurements only. The entangled state of the cluster thus serves as a 
universal substrate for any quantum computation. Cluster states can be created 
efficiently in any system with a quantum Ising-type interaction (at very low 
temperatures) between two-state particles in a lattice configuration.}

We consider two and  three-dimensional arrays of qubits that 
interact via an Ising-type next-neighbor interaction \cite{persistenz-paper} 
described by a Hamiltonian
$H_{\text{int}}= g(t) \sum_{<a,a'>} \frac{1+\sigma_z^{(a)}}{2}\frac{1-\sigma_z^{(a')}}{2} 
\sim -\frac{1}{4}g(t) \sum_{<a,a'>} \sigma_z^{(a)}\sigma_z^{(a')}$ \cite{footnote} 
whose strength $g(t)$ can be controlled externally. A possible realization of such a 
system is discussed below. A qubit at site $a$ can be
in two states $|0\rangle_a$ or $|1\rangle_a$, the eigenstates of the Pauli
phase flip operator $\sigma_z^{(a)}$ ($\sigma_z^{(a)}|i\rangle_a =
{(-1)}^i|i\rangle_a$). These two states form the 
computational basis. Each qubit can equally be in an arbitrary
superposition state
$\alpha |0\rangle + \beta | 1\rangle$, $|\alpha|^2+|\beta|^2=1$. For
our purpose, we initially prepare all qubits in the superposition
$|+\rangle = (|0\rangle + |1\rangle)/\sqrt{2}$, an eigenstate of the
Pauli spin flip operator $\sigma_x$ ($\sigma_x|\pm\rangle = \pm
|\pm\rangle$). $H_{\text{int}}$ is then switched on for an appropriately 
chosen finite time interval, by which a unitary transformation  $S$
is realized. Since $H_{\text{int}}$ acts uniformly on the lattice,
entire clusters of neighboring particles become entangled in one
single step. The quantum state $|\Phi\rangle_{\cal{C}}$, the state of a 
cluster ($\cal C$) of neighboring qubits, which is thereby 
created provides in advance all
entanglement that is involved in the subsequent quantum  
computation. It has been shown \cite{persistenz-paper} \linebreak 
\begin{figure}[tph]
\hspace*{-0.25cm}
 \epsfig{file=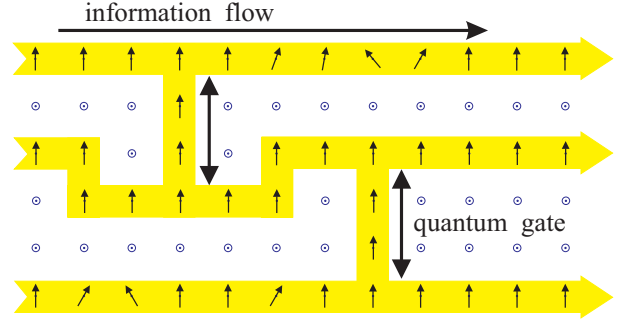,width=8cm}
\vspace*{0.25cm}
\caption{Quantum computation by measuring two-state particles on a
  lattice. Before the measurements the qubits are in the
  cluster state $|\Phi\rangle_{\cal{C}}$ of (\ref{EVeqn}). 
  Circles $\odot$ symbolize measurements of $\sigma_z$, vertical arrows
  are measurements of $\sigma_x$, while tilted arrows refer to
  measurements in the x-y-plane.} 
\label{FIGetching}
\end{figure}
\noindent  that the cluster state $|\Phi\rangle_{\cal C}$ is characterized by a
set of eigenvalue  equations
\begin{equation}
    \sigma_x^{(a)} \prod_{a'\in \text{\tt\normalsize ngbh($a$)}}\sigma_z^{(a')} 
    |\Phi\rangle_{\cal C} = \pm |\Phi\rangle_{\cal C}, 
\label{EVeqn}
\end{equation} 
where {\tt ngbh($a$)} specifies the sites of all qubits 
that interact with the qubit at site $a\in {\cal C}$. The eigenvalues are determined by
the distribution of the qubits on the lattice. The equations (\ref{EVeqn}) are
central for the proposed computation scheme. As an example, a measurement on an 
individual qubit of a cluster has a random outcome. On the other hand, 
(\ref{EVeqn}) imply that any two qubits at sites $a,a^\prime \in
{\cal{C}}$ can be projected into a Bell state by measuring a subset of
the other qubits in the cluster. This property will be used to define 
quantum channels that allow us to propagate quantum information through a cluster.

Here we will show that a cluster state $|\Phi\rangle_{\cal{C}}$ can be used as 
a ``substrate'' on which any quantum circuit can be imprinted by one-qubit measurements. 
In Figure \ref{FIGetching} this scheme is illustrated. For simplicity, we assume that in a
certain region of the lattice each site is occupied  by a qubit. This
requirement is not essential as will be explained below (see d)).
In the
first step of the computation, a subset of qubits is measured in the
basis of $\sigma_z$ which effectively removes them. In
Fig. \ref{FIGetching} these qubits are 
denoted by ``\mbox{$\odot$}''. The state $|\Phi\rangle_{\cal{C}}$ is
thereby projected into a tensor product
$|\mu\rangle_{{\cal{C}}\backslash{\cal{N}}}$ of measured particles on
one side and an entangled network on the other side:
$|\Phi\rangle_{\cal{C}} \longrightarrow
|\mu\rangle_{{\cal{C}}\backslash{\cal{N}}} \otimes
|\tilde{\Phi}\rangle_{\cal{N}}$. The entangled state
$|\tilde{\Phi}\rangle_{\cal{N}}$ is related to the cluster state
$|\Phi\rangle_{\cal{N}}$ on the network ${\cal{N}}$ by a local unitary
transformation which depends on the set of measurement results $\mu$.
$|\tilde{\Phi}\rangle_{\cal{N}}$ still satisfies (\ref{EVeqn}) for all
$a\in{\cal{N}}$ and thus belongs to the same class of entangled states
as the original cluster state $|\Phi\rangle_{\cal{C}}$. The
measurement results $\mu$ enter in the signs in (\ref{EVeqn}) for
$|\tilde{\Phi}\rangle_{\cal{N}}$.

To process quantum information with this network, it suffices to
measure its particles  
in a certain order and in a certain basis. Quantum information is
thereby    
propagated horizontally through the cluster by measuring the qubits on
the wire while 
qubits on  vertical connections are used to realize two-bit quantum
gates. The basis in  which a
certain qubit is  
measured depends in general on the results of the preceding
measurements. The processing is finished once all qubits 
except a last one on  
each wire have been measured. At this point, the results of previous
measurements  
determine in which basis these ``output'' qubits need to be measured for the 
final read-out. 

\begin{figure}[tbp]
\hspace*{-0.25cm}
 \epsfig{file=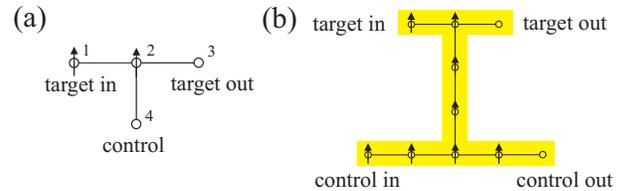,width=8cm}
\vspace*{0.25cm}
\caption{Realization of a CNOT gate by one-particle measurements. See text.}
\label{FIGcnot}
\end{figure}

In the following, we will show that any quantum logic circuit can be
implemented on a cluster state. The purpose of this is twofold. First,
it serves as an illustration of how to implement a
particular quantum circuit in practice. Second, in showing that any quantum
circuit can be implemented on a sufficiently large cluster state we
demonstrate universality of the proposed scheme. For pedagogical
reasons we will first explain a scheme with one 
essential modification with respect to the proposed scheme:  
before the entanglement operation $S$, certain qubits are selected as input 
qubits and the input information is written onto them, while the remaining qubits are
prepared in $|+\rangle$. This step weakens the scheme since it
affects the character of the cluster state as a genuine resource. 
It can, however, be avoided (see e)). Points a) to c) are concerned 
with the basic elements of a quantum circuit, quantum gates and wires, 
point d) with the composition of gates to circuits. 

a) Information propagation in a wire for qubits. A qubit can be teleported
from one site of a cluster to any other site. In particular, consider a chain
of an odd number of qubits 1 to $n$ prepared in the state
$|\psi_{in}\rangle_1 \otimes |+\rangle_2 \otimes ... \otimes |+\rangle_n$
and entangled by \St . The state of qubit 1, $|\psi_{in}\rangle$, can now be
transfered to site $n$ by 
performing $\sigma_x$-measurements (bases
$\{|+\rangle_j=|0\rangle_{x,j},|-\rangle_j=|1\rangle_{x,j}\}$)
at qubit sites $j=1...n-1$ with measurement outcomes $s_j =
\{0,1\}$. The resulting state is $|s_1\rangle_{x,1} \otimes
... \otimes |s_{n-1}\rangle_{x,n-1} \otimes |\psi_{out}
\rangle_n$. The output state $|\psi_{out}\rangle$ is related to the
input state $|\psi_{in}\rangle$ by a unitray transformation $U_\Sigma \in
\{1,\sigma_x,\sigma_z,\sigma_x\sigma_z\}$ which depends on the outcomes of
the $\sigma_x$-measurements at sites 1 to $n-1$. A similar argument
can be given for an even number of qubits. The effect of
$U_\Sigma$ can be corrected for at the end of a computation as will be
shown below (see d)). It is noteworthy that not all
classical information gained by the $\sigma_x$-measurements needs to be
stored to identify the transformation $U_\Sigma$. Instead, $U_\Sigma$ is
determined by the values of only two classical bits which are updated
with every measurement. 

b) An arbitrary rotation $U_R \in SU(2)$ can be achieved in a
chain of 5 qubits. Consider a
rotation in its Euler representation $U_R(\xi,\eta,\zeta) = U_x(\zeta)
U_z(\eta) U_x(\xi)$. The qubits are initially in the  state
$|\psi_{in}\rangle_1 \otimes 
|+\rangle_2 \otimes ... \otimes |+\rangle_5$ and entangled by
\St . Then, qubits 1 to 
4 are measured in the order implied by their numbering. Qubit 1 is
measured in the direction of $\vec{x}$. (Measuring in the direction
of $\vec{r}$ means measuring the operator $\vec{r} \cdot \vec{\sigma}$.) 
The directions for the remaining measurements are all in the
$x$-$y$-plane. Their angles with respect to the $x$-axis are equal to the
respective Euler angles ($\xi,\eta,\zeta$ for qubits 2,3,4) up to a
sign $\pm1$. The signs of the angles  depend on the results of
previous measurements. The final state is then 
$|s_1\rangle_{1,x} \otimes |s_2\rangle_{2,\xi} 
\otimes |s_3\rangle_{3,\eta} \otimes |s_4\rangle_{4,\zeta} \otimes
|\psi_{out}\rangle_5$ with $|\psi_{out}\rangle = U_\Sigma U_R
|\psi_{in}\rangle$. Again, the extra transformation $U_\Sigma \in
\{1,\sigma_x,\sigma_z,\sigma_x\sigma_z\}$ can be read off from the
measurement results and corrected at the end.

c) To perform the gate $\CNOT (c,t_{in}\rightarrow t_{out}) = |0\rangle_c
\,\!_c\langle 0| \otimes 1^{(t_{in} \rightarrow t_{out})} + |1\rangle_c
\,\!_c\langle 1| \otimes \sigma_x^{(t_{in}\rightarrow t_{out})}$
between a control qubit $c$ 
and a target qubit $t$, four qubits, arranged as depicted
Fig.\ \ref{FIGcnot}a, are required. During the action of the gate, the target
qubit $t$ is transfered from $t_{in}$ to $t_{out}$. The following
procedure has to 
be implemented. Be qubit 4 the
control qubit. First, the state $|i_1\rangle_{z,1}
\otimes |i_4\rangle_{z,4}
\otimes  |+\rangle_2 \otimes |+\rangle_3$ is prepared  and then the entanglement
operation $S$ performed. Second, $\sigma_x$ of qubits 1 and 2 is
measured. The measurement results $s_j= \{0,1\}$
correspond to projections of the qubits $j$ into
$|+\rangle_j$ or $|-\rangle_j$, $j=1,2$. The quantum state created by
this procedure is $|s_1\rangle_{x,1} \otimes |s_2\rangle_{x,2} \otimes
U_\Sigma^{(34)}|i_4\rangle_4 \otimes |i_1+i_4 \;\mbox{mod}\;2\rangle_3$, where 
$
     U_\Sigma^{(34)} = {\sigma_z^{(3)}}^{s_1+1} \!\!
     {\sigma_x^{(3)}}^{s_2} \!\!
     {\sigma_z^{(4)}}^{s_1}.
$
The input state is acted upon by the CNOT and successive $\sigma_x$- and
$\sigma_z$-rotations $U_\Sigma^{(34)}$, depending on 
the measurement results $s_1,s_2$. These unwanted extra rotations can
be accounted for as in a) and b). For practical purposes it is more 
convenient if the control qubit is, like the target qubit, transfered to
another site during the action of the gate. When a CNOT is
combined with other gates to form a quantum circuit it will be used in the form
shown in Fig. \ref{FIGcnot}b. 

To explain the working principle of the CNOT-gate we will, for
simplicity, refer to the minimal implementation with four qubits. The
minimal CNOT can be viewed as a wire from qubit 1 to qubit 3 with an
additional qubit, no.\ 4 attached. From the eigenvalue equations
(\ref{EVeqn}) it can now be derived that, if qubit 4 is in a
eigenstate $|i_4\rangle_{z,4}$ of $\sigma_z$ then the value of $i_4 \in
\{0,1\}$ determines whether a unit wire or a spin flip $\sigma_x$ 
(modulo the same correction $U_{\Sigma}^{(3)}$ for both values of $i_4$) 
is being implemented. In other words, once $\sigma_x$ of qubits 1 and 2 
have been measured, the value $i_4$ of qubit 4 controls whether the target 
qubit is flipped or not.    

d) Quantum circuits. The gates described -- the CNOT and arbitrary
one-qubit rotations -- form a universal set \cite{barenco95}.  In the
implementation of a quantum circuit on a cluster state the site of 
every output qubit of a gate overlaps with the site of an input qubit 
of a subsequent gate. This way, the entire entanglement operation can be
performed at the beginning. To see this, compare the following two
strategies. Given a quantum 
circuit implemented on a network ${\cal{N}}$ of qubits which is divided into 
two consecutive circuits. Circuit 1 is implemented on network
${\cal{N}}_1$ and   
circuit 2 is implemented on network  ${\cal{N}}_2$, and ${\cal{N}}={\cal{N}}_1 \cup
{\cal{N}}_2$. There is an overlap 
${\cal{O}}={\cal{N}}_1\cap{\cal{N}}_2$ which contains the sites of the output qubits
of circuit 1 (these are identical to the sites of the input qubits of circuit 2).  
The sites of the readout qubits form a set ${\cal{R}} \subset {\cal{N}}_2$. 
Strategy i) consists of following steps: (1) write input and 
entangle all qubits on ${\cal{N}}$; (2) measure qubits $\in {\cal{N}}\backslash
{\cal{R}}$ to implement the circuit. Strategy ii) consists of (1) write input and 
entangle the qubits on ${\cal{N}}_1$; (2) measure the qubits in
${\cal{N}}_1\backslash {\cal{O}}$. This implements the circuit on ${\cal{N}}_1$ 
and writes the intermediate output to ${\cal{O}}$; (3) entangle the qubits on
${\cal{N}}_2$; (4) measure all qubits in ${\cal{N}}_2 \backslash {\cal{R}}$. 
Step 3 and 4 implement the circuit 2 on ${\cal{N}}_2$.  
The measurements on ${\cal{N}}_1\backslash {\cal{O}}$ commute with the entanglement 
operation restricted to ${\cal{N}}_2$, since they act on different subsets
of particles. Therefore the two strategies are mathematically
equivalent and yield the same results. It is therefore consistent to entangle 
in a single step at the beginning and perform all measurements afterwards.

Two further points should be addressed in connection with
circuits. First, the randomness of the measurement results does not
jeopardize the function of the  circuit. Depending on
the measurement results, extra rotations $\sigma_x$ and $\sigma_z$ act on
the output qubit of every implemented gate. By use of the relations
$\CNOT (c,t) \, \sigma_z^{(t)} = \sigma_z^{(c)} \sigma_z^{(t)} \, \CNOT
(c,t)$, $U_x(\zeta) U_z(\eta) U_x(\xi) \, \sigma_z =   \sigma_z \, U_x(-\zeta)
U_z(\eta) U_x(-\xi)$ and similar ones, these extra rotations can be pulled
through the network to act upon the output state. There they can be
accounted for by adjusting the measurement basis for the final readout.
Rotations require some care. As stated above, changing the
order of a spin flip $\sigma_x$ and an arbitrary rotation reverses the sign of
two Euler angles. This can be compensated for, but introduces a partial temporal
ordering of the measurements on the whole cluster.
Second, quantum circuits can also be implemented on irregular
clusters. In that case, qubits are missing which are
required for the standard implementation of the circuit. This can be
compensated by a large flexibility in shape of the gates and wires. The
components can be bent and stretched to fit to the cluster
structure as long as the topology of the circuit implementation does
not change.  
 
e) Full scheme. It is important to note that the step of writing the input 
information onto the qubits before the cluster is entangled, 
was introduced only for pedagogical reasons. 
For illustration of this point consider a chain of 5 qubits in the 
state $S\, |+\rangle_1 \otimes |+\rangle_2 \otimes ... \otimes |+\rangle_5$.
Clearly, there is no local information on any of the qubits. However, by 
measuring qubits 1 to 4 along suitable directions,
the qubit 5 can be projected into any desired state (modulo $U_{\Sigma}$).
What is used here is the knowledge that the resource has been prepared with qubit 1
in the state $|+\rangle_1$ before the entanglement operation. 
By the four measurements, this qubit is rotated as described in b).
In a similar manner any desired input state can be prepared if the rotations
are replaced by a circuit preceding the proper circuit for computation.
In summary, no input information needs to be written to the qubits before they are 
entangled. Cluster states are thus a genuine resource for quantum computation via 
measurements only.   

A possible implementation of such a quantum computer uses neutral atoms 
stored in periodic micropotentials
\cite{brennen99,jaksch99,calarco99} where Ising-type interactions 
can be realized by controlled collisions between atoms in neighboring 
potential wells \cite{jaksch99,briegel00}. This system combines small decoherence 
rates with a high scalability. 
The question of scalability is, in fact, closely linked to the percolation phenomenon.
For a site occupation probability above the percolation threshold, there exists a cluster
which is bounded  in size only by the trap dimensions. 
For optical lattices in three dimensions, single-atom site occupation with a filling 
factor of 0.44 has been reported \cite{DePue}  which is significantly above the 
percolation treshold of 0.31 \cite{Ziman}. As in other proposed implementations
for quantum computing,  the addressability of single qubits in the lattice 
is however still a problem. (For recent progress see Ref.\ \cite{weitz}).

For a cluster of a given {\em finite} size, the number of computational steps may 
be too large to fit on the cluster. In this case, the  computation can be split 
into several parts, for each of which there is sufficient space on the cluster. 
The modified procedure consists then of repeatedly (re)entangling the cluster and 
imprinting the actual part of the circuit until the whole calculation is performed. 
In this sense, cluster states can be recycled. 
This procedure has also the virtue that qubits involved in the later part 
of a calculation need not be protected from decoherence for a long time while 
the calculation is still being performed at a remote place of the cluster. Standard 
error-correction techniques \cite{shor,steane} may then be used on each part of the 
circuit to stabilize the computation against decoherence.

In conclusion, we have described a new scheme of quantum computation
that consists entirely of one-qubit measurements on a particular class of
entangled states, the cluster states. The measurements are used to
imprint a quantum circuit on the state, thereby destroying its entanglement
at the same time. Cluster states are thus one-way quantum computers and the 
measurements form the program.


\end{multicols}


\begin{thebibliography}{99}
\bibitem{bennett-divincenzo-review}
Bennett, C. H. \& DiVincenzo, D. P. 
Quantum information and computation.
{\em Nature} {\bf 404}, 247--255 (2000).
\bibitem{experiments}
See Ref. \cite{bennett-divincenzo-review} for a recent review.
\bibitem{cirac-zoller-2000}
Cirac, J. I.\& Zoller, P. 
A scalable quantum computer with ions in arrays of microtraps.
{\em Nature} {\bf 404}, 579--581 (2000).
\bibitem{deutsch89}
Deutsch, D. 
Quantum computational networks. 
{\em Proc.\ R. Soc.\ London} {\bf 425}, 
73-90 (1989).
\bibitem{barenco95}
Barenco, A. {\em et al.}
Elementary gates for quantum computation. 
{\em Phys.\ Rev. A} {\bf 52}, 3457--3467 (1995). 
\bibitem{persistenz-paper}
Briegel, H.-J. \& Raussendorf, R. 
Persistent entanglement in arrays 
of interacting particles. Pre-print quant-ph/0004051 at 
$\langle$xxx.lanl.gov$\rangle$ (2000).
\bibitem{footnote}
The second Hamiltonian is of the standard Ising form. The symbol $\sim$
means that the states generated from a given initial state, under the 
action of these Hamiltonians, are identical up to a local rotation on 
certain qubits. We use the first Hamiltonian to make the computational 
scheme more transparent. The conclusions drawn in the paper are, however,
the same for both Hamiltonians. 
\bibitem{brennen99}
Brennen, G. K., Caves, C. M., Jessen, P. S. \& Deutsch, I. H. 
Quantum Logic Gates in Optical Lattices.  
{\em Phys.\ Rev.\ Lett.} {\bf 82}, 
1060-1063 (1999).
\bibitem{jaksch99}
Jaksch, D., Briegel, H.-J., Cirac, J. I., Gardiner, C. W. \& Zoller, P.  
Entanglement of Atoms via Cold Controlled Collisions.  
{\em Phys.\ Rev.\ Lett.} {\bf 82}, 1975-1978 (1999).
\bibitem{calarco99} 
Calarco, T., Jaksch, D., Cirac, J. I. \& Zoller, P. 
Quantum Gates with neutral atoms: Controlling collisional interactions 
in time dependent traps. 
{\em Phys.\ Rev.\ A} {\bf 61}, 022304 (2000). 
\bibitem{briegel00}
H.-J. Briegel, Calarco, T., Jaksch, D., Cirac, J. I. \& Zoller, P. 
Quantum computing with neutral atoms. 
{\em J.\ Mod.\ Opt.} {\bf 47}, 415-451 (2000).
\bibitem{DePue}
DePue, M. T., McCormick, C., Winoto, S. L., Oliver, S. \& Weiss, D. W. 
Unity Occupation of Sites in a 3D Optical Lattice, 
{\em Phys.\ Rev.\ Lett.} {\bf 82}, 2262-2265 (1999).
\bibitem{Ziman}
Ziman, J. M. Models of Disorder. Cambridge University Press, 1979.
\bibitem{weitz}
Scheunemann, R., Cataliotti, F. S., H\"ansch, T. W, \& Weitz M. 
Resolving and addressing atoms in individual sites of a 
CO$_2$-laser optical lattice. Accepted by Phys.\ Rev.\ A. 
\bibitem{shor}
Calderbank, A. \& Shor, P. W. 
Good quantum error correcting codes exist.
{\em Phys.\ Rev.\ A} {\bf 54}, 1098 (1996).
\bibitem{steane}
Steane, A. 
Error Correcting Codes in Quantum Theory.
{\em Phys.\ Rev.\ Lett.} {\bf 77}, 793 (1996).

\end{thebibliography}
\end{document}